\documentclass{aastex}
 \usepackage{emulateapj5}
 \usepackage{apjfonts}

\slugcomment{KUNS-1629}

\shorttitle{TASHIRO \& NISHI}
\shortauthors{THERMAL INSTABILITY AND CLUSTER FORMATION}

\begin{document}            

\title{ON THE THERMAL INSTABILITY IN A CONTRACTING GAS CLOUD AND FORMATION OF
A BOUND CLUSTER}

\author{MOTOMICHI TASHIRO AND RYOICHI NISHI}
\affil{Department of Physics, Kyoto University,
    Kyoto 606-8502, Japan}

\affil{tashiro@tap.scphys.kyoto-u.ac.jp, nishi@tap.scphys.kyoto-u.ac.jp}

\begin{abstract}
We perform linear analysis of thermal instability in a contracting large 
cloud filled with warm \ion{H}{1} gas and investigate the effect of 
metallicity and radiation flux. When the cloud reaches critical density 
$n_{f}$, the cloud fragments into cool, dense condensations because of 
thermal instability. For a lower metallicity gas cloud, the value of $n_{f}$ 
is high. Collision between condensations will produce self-gravitating clumps 
and stars thereafter. From the result of calculation, we suggest that high 
star formation efficiency and bound cluster formation are realized in 
low-metallicity and/or strong-radiation environments.  
\end{abstract}


\section{Introduction}

The Galactic globular clusters are suggested to be fossil record of 
the Galaxy formation era (e.g., Fall $\&$ Rees 1985). 
So any knowledge about their formation mechanism will be helpful 
to understand the protogalactic environment.    
Young counterparts of globular clusters are absent in the Milky Way, 
whereas in galaxies such as the LMC, SMC and M33, young globulars exist 
(e.g., Kumai, Basu, $\&$ Fujimoto 1993; Efremov 1995).
Such galaxies are in many case ongoing mergers and merger remnants 
or starburst galaxies, but young globulars also exist in normal spiral, 
dwarf and irregular galaxies (Schweizer 1999). 
The conditions that enable globular cluster formation in these galaxies 
may be also realized at the epoch of the Galactic globular cluster formation. 
Since globular clusters are gravitationally bound, 
we attempt to clarify the condition of bound cluster formation.   

At the formation epoch of the globular clusters, high star formation 
efficiency (SFE) must be achieved somehow, since high SFE is required in 
order to form a bound cluster. The threshold SFE value for bound cluster 
formation is 0.5 in the case of rapid gas removal (e.g., Hills 1980), which is 
significantly greater than average SFE ($\sim$ 0.01) in the giant molecular 
clouds (GMCs).
Thus, to understand the conditions of bound cluster formation, 
we should know in what environment the SFE can be high.
In regard of this problem, Elmegreen $\&$ Efremov (1997) claimed that 
high-pressure environment is required for bound cluster formation 
because disruption of the cluster-forming cloud by massive stars is difficult 
in such an environment. 
They also argued that bound cluster formation is efficient in a 
low-metallicity environment, as it is difficult to disrupt the cloud by 
radiation pressure of massive stars. 
Here, considering the formation process of dense clouds, we point out that 
high SFE and efficient bound cluster formation can be 
achieved in low-metallicity and/or strong-radiation environments because of 
cloud fragmentation via thermal instability.  

Molecular clouds, where stellar clusters form, usually belong to some 
larger structure such as superclouds (Elmegreen $\&$ Elmegreen 1983), 
the largest cloud complexes in galaxies. 
Due to the similarity between the masses of superclouds and 
the critically unstable Jeans mass,
large-scale gravitational instability is considered to produce 
these superclouds (Elmegreen 1987). 
Warm \ion{H}{1} gas occupies large fraction of these superclouds, 
since roughly half of the \ion{H}{1} in the Milky Way disk 
is in the form of warm gas (Kulkarni $\&$ Heiles 1988).
When the gas disk becomes gravitationally unstable and fragments into 
superclouds, this warm gas should be converted into cold \ion{H}{1} gas 
since the gas becomes thermally unstable when the density is somewhat higher. 
From this cold \ion{H}{1} gas, GMCs and, subsequently, stellar clusters are 
expected to form.
Similarly, GMC formation in a supercloud via thermal instability was also 
considered by Kolesnik (1991). However, his method was based on density 
structure of static isothermal cloud and our treatment is different.

For modeling the above-mentioned top-down scenario of cluster formation from 
a supercloud (see also Efremov 1995), we consider a large, warm gas cloud 
($T \sim 10^{4} {\rm K} $) contracting because of its self-gravity. 
As the cloud density increases, it becomes thermally unstable and 
breaks up into cool, dense condensations. 
We investigate this process based on linear analysis of thermal instability 
in the cloud. For the evolution of condensations, we adopt the scenario that 
these condensations collide each other and 
self-gravitating clumps are formed in the cloud, 
and then stellar clusters form. To judge whether the cluster is bound or not, 
we should know the mean density of the cloud at the point of breakup,
since we can relate the mean density of the cloud to its velocity dispersion, 
which determines SFE and hence whether the cluster is bound or unbound. 
So we calculate the thermal evolution of a contracting gas cloud and 
estimate the density when the cloud becomes thermally unstable. 
The effects of metallicity and radiation flux are also examined. 

In the following sections, we present the model of the cloud and 
the instability analysis method. Based on the results of the calculations,
we discuss the conditions necessary for high SFE and bound cluster formation. 

\section{Thermal Evolution of a Contracting Gas Cloud} 

We calculate the evolution of density, temperature, and ionization degree of 
a contracting gas cloud. For simplicity, we consider a spherical and 
homogeneous gas cloud. As an initial condition, we take the static cloud 
filled with warm \ion{H}{1} gas ($T \sim 10^{4} {\rm K} $) in thermal and 
ionization equilibrium. The contraction is simplified as a free-fall collapse.
Then the thermal evolution of the cloud is given by the following 
basic equations:
\begin{equation}
  \frac{d}{dt} \ln n = \left\{ 24 \pi G \rho \left[ 1 - \left( \frac{n_{i}}
  {n} \right)^{1/3} \right] \right\}^{1/2} , 
\end{equation}
\begin{equation}
  \frac{d \varepsilon}{dt} = \frac{P}{ \rho^{2}} \frac{d \rho}{dt}
 - {\cal L} ,
\end{equation}
\begin{equation}
 \frac{d x_{e}}{dt} = X \left( T,\rho,x_{e} \right) .
\end{equation}
Here $\rho$ and $n$ are the mass and number density of hydrogen of 
the cloud, respectively, and $n_{i}$ is the initial value of $n$. 
$P$, $T$, $\varepsilon$, and $x_{e}$ are the pressure, temperature, 
internal energy, and ionization degree of the cloud, respectively. 
Function ${\cal L}$ represents the net cooling rate.
As heating sources, we consider interstellar UV radiation, soft X-ray, 
and cosmic ray. 
As coolants, we consider hydrogen ${\rm Ly}\alpha $ cooling, line cooling by 
\ion{O}{1}, \ion{C}{2}, \ion{Si}{2}, \ion{Fe}{2}, dust cooling, and so on. 
Ionization states of these metal species 
are fixed according to their ionization potentials. 
In this work we approximate the cloud as optically thin. 
Function $X$ represents the net ionization rate of hydrogen atoms.
We consider soft X-ray and cosmic ray as ionizing sources. 
To determine the X-ray heating and ionization rate, the hydrogen column 
density of the cloud, $N_{\rm H}$, must be specified. 
we take $N_{\rm H}=10^{19}{\rm cm}^{-3}$ throughout the cloud evolution, 
though $N_{\rm H}$ increases as the cloud contracts. 
This simplification is not serious, since 
we performed the same calculation including the change of $N_{\rm H}$ and 
got almost the same result. 
For these thermal processes, we refer to Wolfire et al. (1995) and references 
there in.

To verify the dependence on metallicity and radiation flux,
we introduce parameters $Z$ and $X_{0}$.
In our assumptions, the number density of dust grains and the gas phase 
abundances of \ion{O}{1}, \ion{C}{2}, etc. change in proportion to $Z$. 
The UV radiation flux, X-ray flux and cosmic ray flux change 
proportional to $X_{0}$. 
Here $Z = 1$ and $X_{0} = 1$ correspond to solar values.  

\section{Formation of Condensations Due to Thermal Instability}

We estimate the mass and the formation timescale of condensations
using linear approximation. The growth of density perturbation in the parent 
cloud is governed by the equation 
\begin{eqnarray}
   \sigma^3 + \sigma^2 c_s \left( k_T + \frac{k^2}{k_K} \right) 
   + \sigma {c_s}^2 k^2 + \frac{ {c_s}^3 k^2 }{\gamma} \nonumber \\
   {} \times \left[ k_T - k_{\rho} + \frac{k^2}{k_K}
   - \frac{\gamma}{{c_s}^3} \left(\gamma -1 \right) {\cal L}_{0} \right] 
   = 0,
\end{eqnarray}
where $k$ and $\sigma$ are the wave number and the growth rate of the 
perturbation, respectively.   
This equation is the same as in Field (1965), 
except in the last term, where it represents nonequilibrium effect. 
Thus, the net cooling rate 
${\cal L}_{0} = {\cal L} \left( T_{0}, \rho_{0}, x_{e0} \right)$ 
is not necessary zero, where $T_{0}$, $\rho_{0}$, and $x_{e0}$ are 
the background temperature, density, and ionization degree, respectively. 
The definitions of $k_{\rho}$, $k_T$, and $k_K $ are the same as 
in Field (1965):  
\begin{eqnarray}
  k_{\rho} = \frac{ \mu \left( \gamma -1 \right) \rho_0 {\cal L}_{\rho}
  }{ R c_s T_0 } ~,~
  k_T = \frac{ \mu \left( \gamma -1 \right) {\cal L}_T }{ R c_s } ~,~  
  \nonumber \\
  {} k_K = \frac{ R c_s \rho_0 }{ \mu \left( \gamma -1 \right) K_0 },
  ~ ~ ~ ~ ~ ~ ~ ~ ~ ~ ~ ~ ~ ~  
\end{eqnarray} 
where $\mu$, $\gamma$, $R$, and $c_s$ are the mean molecular weight, 
ratio of specific heats, gas constant, and sound speed, respectively. 
We take the value of $\mu$ as $1.4$ and $\gamma$ as 5/3. 
The sound speed is expressed as  
$c_s = \left( \gamma P_0/\rho_0 \right)^{1/2}$, where 
$P_{0}$ is the background pressure. 
The symbols ${\cal L}_{\rho}$, ${\cal L}_{T}$, and $K_{0}$ are represented 
as follows: 
\begin{equation}
 {\cal L}_{\rho} = \left( \frac{\partial {\cal L}}{\partial
 \rho } \right)_T ~,~ {\cal L}_{T} = \left( \frac{\partial {\cal L}}
  {\partial T} \right)_{\rho} ~,~ K_{0} = K \left( T_{0} \right),
\end{equation}
where $K(T)$ is thermal conductivity, and we adopt $K(T) = 2.0 \times 10^{3} 
~ T^{1/2}$, the thermal conductivity for neutral hydrogen.

Among three roots of equation (4), two roots correspond to sound wave modes 
and one root corresponds to isobaric condensation mode (Field 1965). 
When this equation has unstable isobaric condensation mode,
condensations form because of the growth of unstable perturbations 
(e.g., Goldsmith 1970).
This mode has a maximum growth rate of $\sigma_{m}$ at 
wave number $k_{m}$ (Fig. 1). 
Among various density perturbations with different intensities and wave 
numbers in the cloud, the perturbations with wave number $k_{m}$ grow 
fastest. So we estimate the mass of the condensation, $M$, as
\begin{equation}
 M = \frac{4\pi}{3} \rho_{0} \left( \frac{\lambda_{m}}{2} \right)^{3}, 
~ ~ \lambda_{m} = \frac{2 \pi}{k_{m}}.
\end{equation}
We also estimate the formation timescale as
\begin{equation}
t_{g} = \frac{1}{\sigma_{m}}.
\end{equation} 

\begin{figure*}
\figurenum{1}
\epsscale{0.8}
\plotone{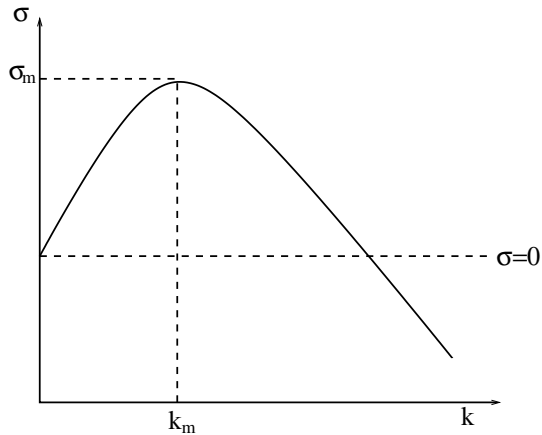}
\figcaption{Dispersion relation of thermal instability 
(schematic example). Among three modes, only condensation mode is 
presented. Since the region $\sigma > 0$ exists, this mode is unstable and 
condensations are formed from density perturbations with positive $\sigma$. 
Note that there exists a maximum growth rate $\sigma_{m}$ at wave number 
$k_{m}$.}
\end{figure*}

Because equation (4) is derived under the assumption that 
contraction of the cloud is negligible, it must keep in mind that 
equation (4) is applicable only when the growth timescale of perturbation is 
shorter than the contraction timescale of the cloud. 
Also note that our use of equation (4) is justified when $\lambda_{m}$ 
is much smaller than the parent cloud, because equation (4) is for 
an infinite homogeneous medium. 
We can neglect this point, since the parent cloud ($\ga$ a few kpc ) 
is much larger than $\lambda_{m}$ ( $\sim$ a few pc).

\section{Results}

We perform the calculation in the parameter ranges $Z= 1$-$10^{-2}$, 
$X_{0} = 10$-$0.1$.
For the purpose of illustration, we show the results of the cases
($Z=1,X_{0}=1$) and ($Z=10^{-2},X_{0}=1$) in Figures 2 and 3. 
In the figures, evolutionary paths in $n-T$ plane of both the free-fall 
collapsing gas cloud and the quasi-statically contracting gas cloud keeping 
thermal equilibrium are presented. 
Equations (1)-(3) are used for the free-fall collapsing paths 
and the equations, ${\cal L} \left(T,\rho,x_{e} \right) = 0$ and 
$X \left(T,\rho,x_{e} \right)=0$, are used for thermal equilibrium paths. 
The condensation mass $M$ and $t_{g}/t_{dyn}$, the ratio of
formation timescale of condensation to dynamical timescale of the parent 
cloud, are also shown in the same figures. 
Here the dynamical timescale of the parent cloud, $t_{\rm dyn}$, is defined as 
$\rho / \left(d \rho /dt \right)$, where $\rho$ is the density of 
the parent cloud. The contours of $M$ and $t_{\rm g}/t_{\rm dyn}$ 
in Figures 2 and 3 are calculated using equation (4), with the assumption of 
ionization equilibrium (i.e., only $n$ and $T$ are used for calculation, 
and $x_{e}$ is a function of $n$ and $T$). 
From the figures we can see the tendency that the smaller metallicity,
the more the evolutionary path of the free-fall collapsing cloud deviates 
from the path of the quasi-statically contracting cloud. 
For the case $Z=1$, the two evolutionary paths deviate in temperature 
$\sim 1000 {\rm K}$ at density $\sim 1 ~{\rm cm}^{-3}$. 
In contrast, for the case $Z = 10^{-2}$, the two paths deviate 
$\sim 8000 {\rm K}$ at density $\sim 100 ~{\rm cm}^{-3}$.  
For the paths of the free-fall collapsing gas cloud, 
dependence on metallicity is stronger than equilibrium paths. 
Such behavior is explained by the fact that a nonequilibrium state is 
realized for free-fall cases because the cooling timescale can be long 
relative to dynamical time and that degree of nonequilibrium is stronger for 
lower metallicity since cooling time is in inverse proportion to $Z$. 

As the cloud density exceeds some critical value, the gas 
cloud becomes thermally unstable. At this point, the growth timescale of 
perturbation, $t_{g}$, is longer than the dynamical timescale of the cloud, 
$t_{\rm dyn}$. So the perturbation cannot grow to the condensation, because 
the parent cloud evolves faster than the perturbation. As the cloud evolves, 
$t_{g}$ becomes smaller than $t_{\rm dyn}$. 
Thus, condensation will form at the moment $t_{g} = t_{\rm dyn}$.
For the case $Z=1$, condensations form at the density 
$\sim 1 ~{\rm cm}^{-3}$ (Fig. 2a), whereas for the case $Z=10^{-2}$, 
condensations form at the density $\sim 300 ~{\rm cm}^{-3}$ (Fig. 3a). 
Condensation mass is a few $10 M_{\odot}$ at the density 
$\sim 1 ~{\rm cm}^{-3}$ for $Z=1$ (Fig. 2b) and a few $M_{\odot}$ at the 
density $\sim 300 ~{\rm cm}^{-3}$ for $Z=10^{-2}$ (Fig 3b).

\begin{figure*}
\figurenum{2}
\epsscale{2.2}
\plottwo{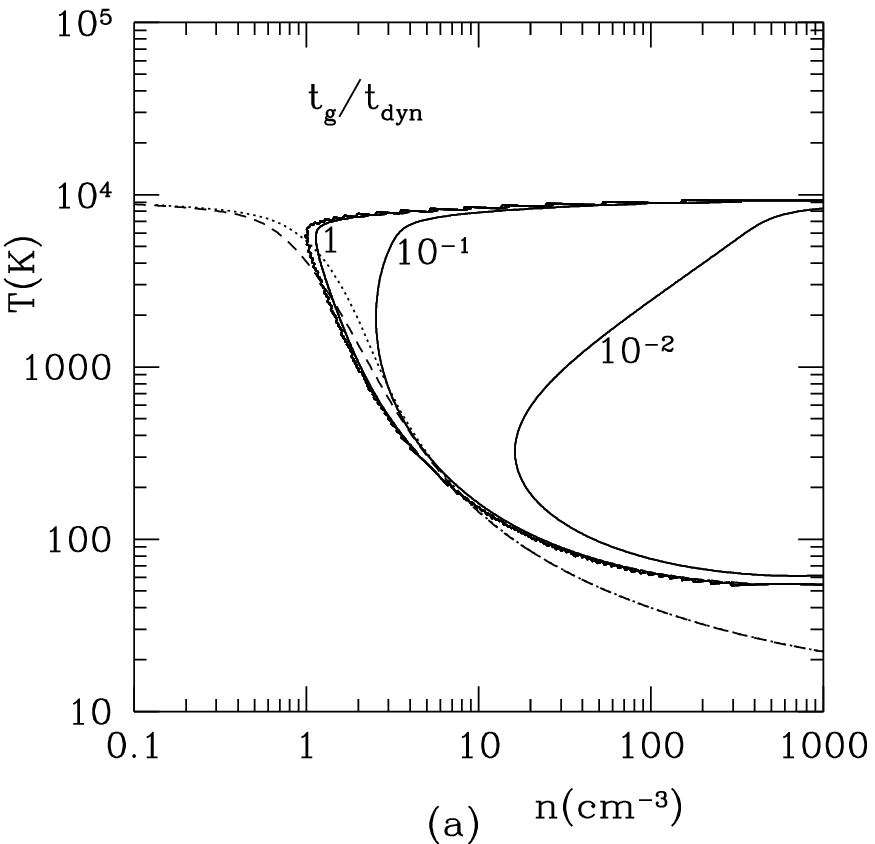}{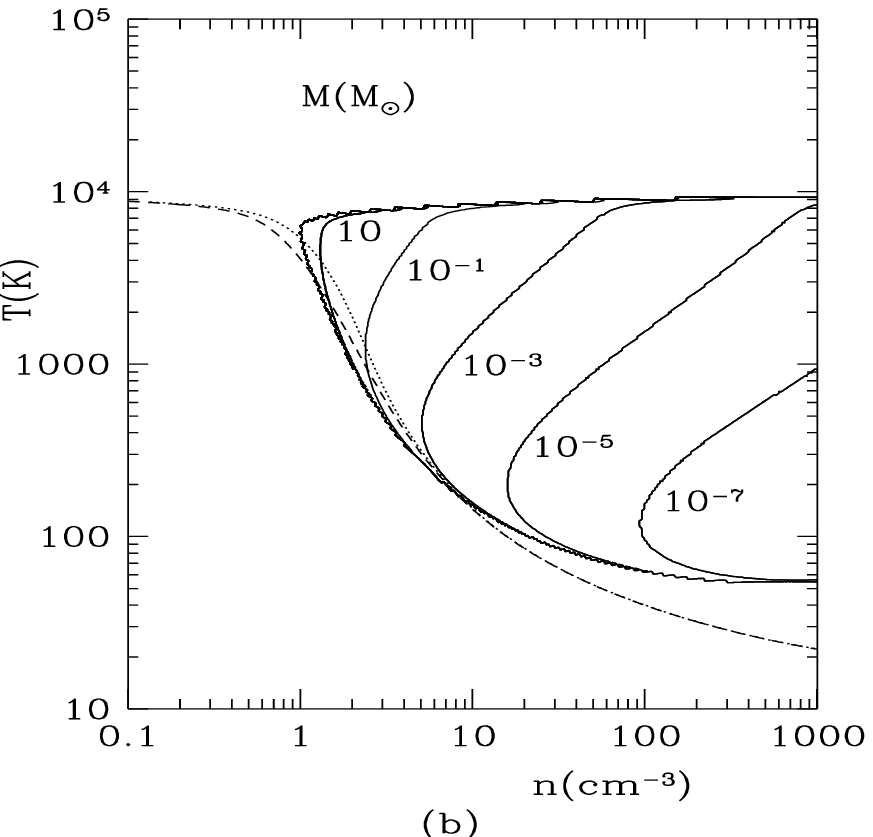}
\caption{
(a) Ratio of formation timescale of condensation, $t_{g}$, 
to dynamical timescale of the cloud, $t_{\rm dyn}$, is shown in $n-T$ plane 
as solid contours. The dynamical time $t_{\rm dyn}$ is defined as 
$\rho_{0}/ \left(d \rho_{0}/dt \right)$, where $\rho_{0}$ is the density of 
the cloud. Numbers near the contours represent the values of 
$t_{g}/t_{\rm dyn}$. Metallicity and radiation parameters are $Z=1$ and 
$X_{0}=1$, respectively.
The thermally unstable region is inside the most outer contour. 
We calculate the contours under the assumption of ionization equilibrium.
Thermal evolution of the free-falling cloud is presented in the plane as 
a dotted line. For reference, we also show thermal evolution of 
the quasi-statically contracting cloud keeping thermal equilibrium 
as a dashed line. 
(b) Condensation mass $M(M_{\odot})$ is shown in 
$n-T$ plane as solid contours for the same parameters. 
Numbers near the contours represent the values of $M(M_{\odot})$.
Meanings of dashed and dotted lines are the same as in Fig. 2a.
}
\end{figure*}

\begin{figure*}
\figurenum{3}
\epsscale{2.2}
\plottwo{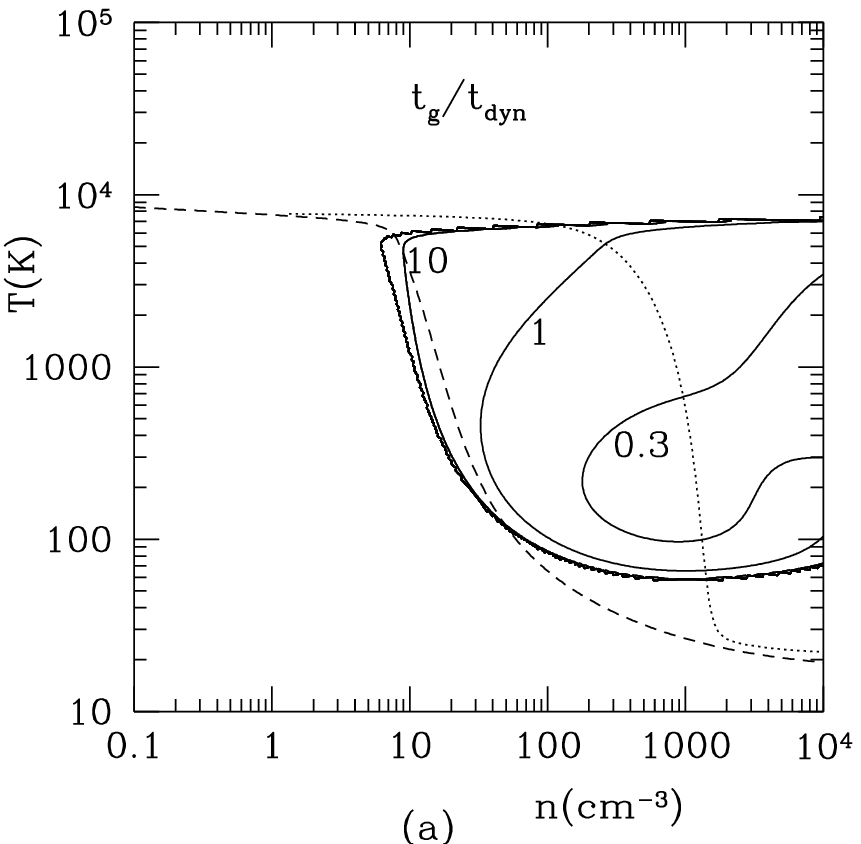}{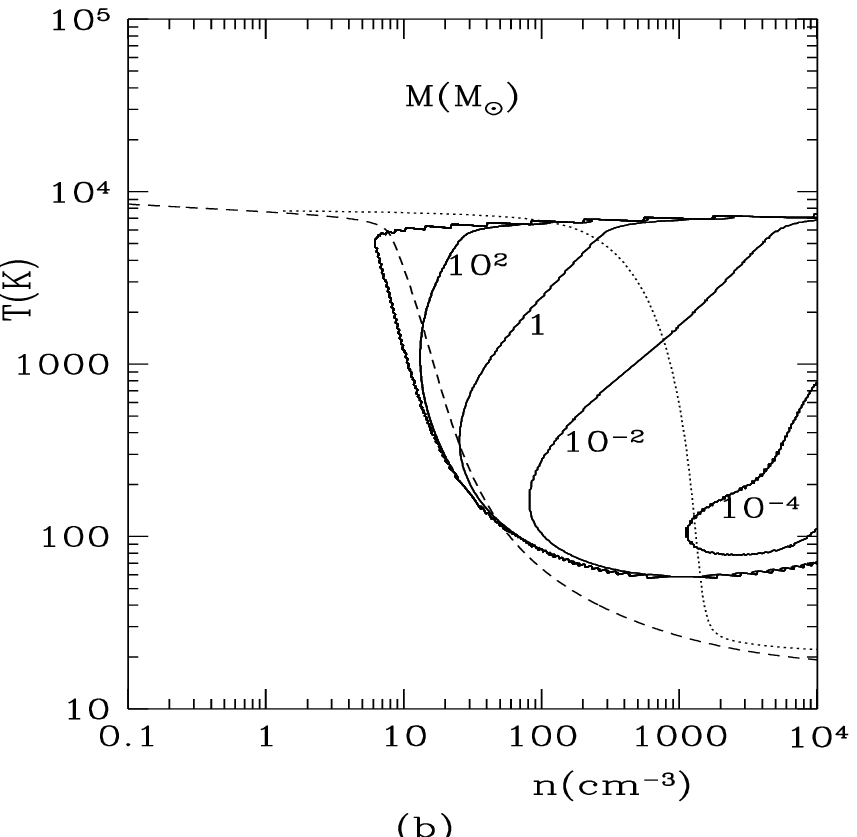}
\caption{Same as Fig. 2, but the metallicity parameter is $Z=10^{-2}$.
The evolutionary path of free-fall collapse case deviates far from the path 
of quasi-statically contracting case.
}
\end{figure*}

For the case when $Z$ and $X_{0}$ are simultaneously changed, 
the results are shown in Figure 4. 
The calculational procedure for Figure 4 differs from Figures 2 and 3, in a 
point that $t_{g}$ and $M$ are calculated using equations (1)-(8) to 
draw Figure 4, whereas the assumption of ionization equilibrium is adopted to 
draw the contours of Figures 2 and 3. 
In Figure 4a, the cloud density $n_{f}$ at the time of condensation formation 
is shown. For $X_{0}=1$, the density $n_{f}$ increases from 
$\sim 1 {\rm cm}^{-3}$ at $Z =1$ to $\sim 300 {\rm cm}^{-3}$ at $Z = 10^{-2}$. 
For $Z = 1$, $n_{f}$ increases from $\sim 0.1 {\rm cm}^{-3}$ at $X_{0} = 0.1$ 
to $\sim 10 {\rm cm}^{-3}$ at $X_{0} = 10$.     
In Figure 4b, the condensation mass is shown. 
For $X_{0}=1$, the mass $M$ decreases from $\sim 30 M_{\odot}$ at $Z =1$ 
to $\sim 1 M_{\odot}$ at $Z = 10^{-2}$. 
For $Z = 1$, $M$ decreases from $\sim 300 M_{\odot}$ at $X_{0} = 0.1$ to 
$\sim 1 M_{\odot}$ at $X_{0} = 10$.   
Thus, from Figure 4, we can see that for the smaller metallicity, 
the cloud density $n_{f}$ is the larger, and 
the condensation mass $M$ is the smaller.
We can also see that for the stronger radiation, 
the density $n_{f}$ is the larger and the mass $M$ is the smaller.

\begin{figure*}
\figurenum{4}
\epsscale{2.2}
\plottwo{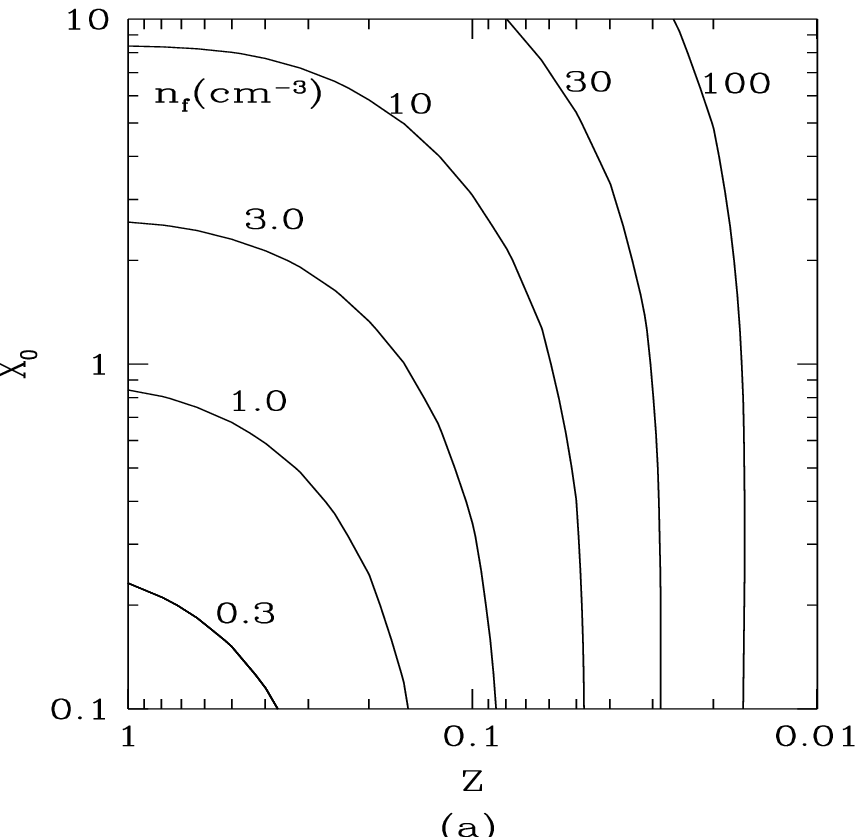}{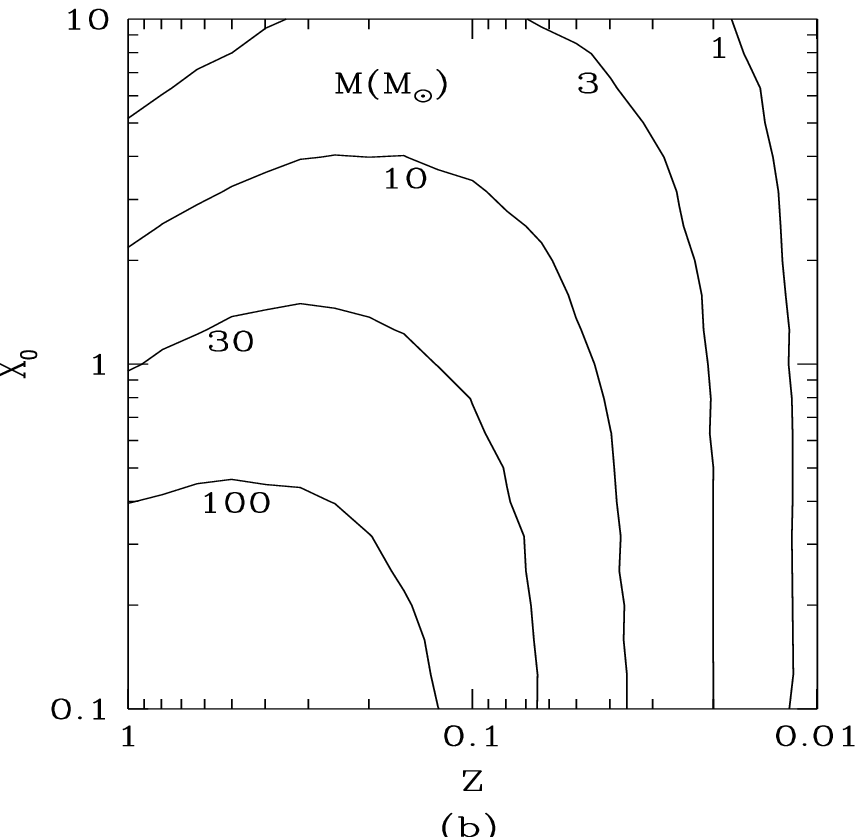}
\caption{(a) Cloud density $n_{f}({\rm cm}^{-3})$ at 
the time of condensation formation is presented in $X_{0}$-$Z$ 
plane as contours. 
Numbers near the contours represent $n_{f}({\rm cm}^{-3})$.
We assume that condensations form at the moment $t_{g} = t_{\rm dyn}$, 
where $t_{g}$ is the formation timescale of condensation and $t_{\rm dyn}$ 
is the contraction timescale of the cloud.  
(b) Condensation mass $M(M_{\odot})$ at the time of 
its formation is presented in $X_{0}$-$Z$ plane as contours.
Numbers near the contours represent $M(M_{\odot})$.}
\end{figure*}

Previously, Elmegreen \& Parravano (1994) calculated 
the critical pressure for coexistence of cool and warm gas, changing 
radiation flux and metallicity. This critical pressure roughly 
corresponds to $n_{f}$ in this work. 
Contrary to our results, metallicity dependence of the critical pressure 
is small in their calculation. 
They assumed static cloud for the background, whereas we assume 
free-falling cloud. 
Then, as noted above, for the smaller metallicity,
the evolutionary path of free-fall contracting cloud deviates more from 
the path of the quasi-statically contracting cloud, because of the reduced 
cooling rate. 
On the other hand, for the static cloud, thermal equilibrium is achieved 
so that metallicity dependence is weak. 
This is the reason of the difference between our work and their work.

We estimate the density of condensation as follows.
After the criterion $t_{g} = t_{\rm dyn}$ is achieved,  
the overdensity region (size $\sim$ $\lambda_{m}$) in the parent cloud 
grows in pressure equilibrium with surrounding gas, since crossing time is 
short compared with cooling time for the fastest growing perturbation. 
Then the overdensity evolves along an isobaric path starting from the point 
$t_{g} = t_{\rm dyn}$ (this path is not drawn in Figs.2 and 3). 
The runaway growth of the overdensity continues until it reaches 
thermochemical equilibrium, where it is thermally stable 
(e.g., Parravano 1987). 
We search the intersection of the isobaric path (the path with 
$nT={\rm constant}$, starting from the point $t_{g} = t_{\rm dyn}$) 
with the equilibrium path (Figs.2 and 3, {\it dashed line}) 
and identify it as ``condensation''.
From the intersection, we can know the density of condensation, $n_{c}$, 
and the temperature of the condensation, $T_{c}$. 
We perform the calculation and find that $n_{c}$ is roughly 100 times larger 
than $n_{f}$ in most parameter ranges. 
In the above-described procedure, we assume that the intersection is 
thermally stable and self-gravity of the overdensity can be neglected.
We confirm that all intersections in this calculation are thermally stable,  
so that the first assumption is justified.
The second assumption is also justified by the later discussion. 
Using the values of $n_{c}$ and $T_{c}$, the Jeans mass can be derived, 
and it is much larger than the condensation mass for most parameter values. 
Thus, self-gravity of the condensation is negligible and 
the density inside the condensation is expected to be almost uniform. 
After formation, condensations accrete surrounding gas and can potentially 
grow to be self-gravitating. 
To estimate the mass growth rate of condensation, we should calculate the 
rate of mass flow across the boundary between the condensation and 
surrounding gas. According to previous works 
(e.g., Penston \& Brown 1970; Parravano 1987), the growth timescale of the 
condensation mass is much larger than the free-fall 
timescale of the parent cloud. So we can neglect the mass growth 
of condensation through accretion. 

\section{Application to bound cluster formation}

After formation of condensations, the parent cloud continues 
free-fall contraction until collisions between condensations begin. 
Then the formation of massive self-gravitating clumps through collisional 
buildup of small condensations is expected. Star formation inside such 
massive clumps will follow. However, evaluation of such evolution is 
difficult task. The onset of collisions depends on velocity distribution 
and space density of condensations, which we cannot know precisely. 
Subsequent evolution of condensations also depends on velocity and 
mass distribution of condensations through the collisional process. 
This collisional process is not so simple (e.g., Hausman 1981) and makes 
the calculation of evolution complicated. 
In regard of these difficulties, we adopt a simplified picture of 
the evolution of condensations. 
We neglect the random velocity of condensations and assume homologous 
contraction of the parent cloud composed of condensations. 
Then the collisions begin when the mean parent cloud density increases to 
the density of condensation, $n_{c}$, because the parent cloud is almost 
filled with condensations in such situation.
As a result of collisions (we assume that the output of collision is mostly 
adhesion), massive self-gravitating clumps will form. 
At this point, the density of massive clumps is larger than 
the mean parent cloud density $n_{c}$, because of self-gravity.
So the parent cloud contract further. 
We assume collisionless collapse of the system composed of 
self-gravitating clumps, for simplicity. 
Then the virialization of the system and the beginning of star formation 
in the clumps is expected. 

It is reasonable to consider that the cluster-forming region is 
only a part of the parent cloud, not the entire cloud. 
We define the mass of the region as $M_{\rm cl}$ and take the reference value
as $10^{5} M_{\odot}$ according to typical mass of globular clusters. 
Then virial velocity of the region is 
\begin{equation}
c \sim 2 \times \left( \frac{M_{\rm cl}}{10^{5}M_{\odot}} \right)^{1/3}
 \left( \frac{n_{\rm cl}}{1 {\rm cm}^{-3}} \right)^{1/6} ~ 
{\rm km} ~{\rm s}^{-1},
\end{equation} 
where $n_{\rm cl}$ is the mean hydrogen number density of the 
cluster-forming region. 
Because (1) we assume collisionless collapse of the parent cloud 
composed of self-gravitating clumps, where the mean cloud density at the 
onset of collapse is the same as condensation density $n_{c}$, 
and (2) condensation density $n_{c}$ is typically 100 times higher than 
the parent cloud density at the time of condensation formation, $n_{f}$ 
(see the last paragraph in {\S} 4), $n_{\rm cl}$ and $n_{f}$ are related 
as
\begin{equation}
n_{\rm cl} \sim 800 n_{f},
\end{equation}
where the relations, $n_{\rm cl} \sim 8 n_{c}$ from the point (1) and 
$n_{c} \sim 100 n_{f}$ from the point (2), are used.

Whether the cluster is bound or not depends on SFE, 
the ratio of stellar mass to the mass of cluster-forming region.
And SFE is determined by the interaction between stellar activity and 
the cluster-forming region. For example, stellar wind, radiation pressure, 
ionization pressure of the \ion{H}{2} region or supernova explosion 
may blow off remaining gas and lead to lower SFE. In this paper we consider 
two cases, gas removal by the expanding \ion{H}{2} region and supernova 
explosion. 

First we discuss gas removal by expanding \ion{H}{2} region. 
The pressure of \ion{H}{2} region corresponds to velocity dispersion 
$c_{\rm HII} \sim 10 {\rm km}~{\rm s}^{-1}$. 
If virial velocity of the region, $c$, is smaller than $c_{\rm HII}$, 
the \ion{H}{2} region pushes out the surrounding gas and blows it off, 
and then the cluster will be unbound. 
In opposite case, the \ion{H}{2} region cannot blow off the surrounding gas 
and high SFE can be achieved; then the cluster will be bound. 
So the criterion of bound cluster formation is considered as $c > c_{\rm HII}$.
This corresponds to the condition
\begin{equation}
n_{f} > 10 \times \left( \frac{C}{800}
\right)^{-1} \left( \frac{c_{\rm HII}}{10{\rm km}~{\rm s}^{-1}} \right)^{6} 
\left( \frac{M_{\rm cl}}{10^{5}M_{\odot}} \right)^{-2} {\rm cm}^{-3},
\end{equation}
where $C$ represents density enhancement, i.e., the ratio of $n_{\rm cl}$ to 
$n_{f}$. 
Note that derivation of equation (11) relies on some rough 
assumptions and depends on uncertain factors such as $C$, $c_{\rm HII}$, and  
$M_{\rm cl}$. Moreover, dependence on $c_{\rm HII}$ is 
very strong. So we should regard equation (11) as qualitative criterion. 

Next we discuss gas removal by supernova (SN) explosion. 
The proto-globular cluster cloud is disrupted by 
several to tens of concurrent 
SN explosions (Dopita \& Smith 1986; Morgan \& Lake 1989).
A high star formation rate at the formation epoch of globular cluster 
will cause such multiple SN explosions so that relic gas will be blown off. 
When the star formation time scale $t_{\rm SF}$ is shorter than the lifetime 
of massive star, $\tau$, a large fraction of gas in the cluster-forming region 
can be converted into stars before SN explosions; then high SFE is 
achieved and a bound cluster can be formed.
In contrast, when $t_{\rm SF}$ is longer than $\tau$, 
SN explosions will remove surrounding gas and stop later star formation.
In this case, low SFE and unbound cluster formation is expected 
(e.g., Yoshii and Arimoto 1987).
The star formation timescale $t_{\rm SF}$ is expected as 
\begin{equation}
 t_{\rm SF} = A t_{ff}, 
\end{equation}
where $A$ represents a proportional constant of order 1 and
$t_{ff} = \left( 3 \pi / 32 G \rho \right)^{1/2}$ is free-fall timescale of 
the cluster-forming region. 
We adopt the lifetime of massive star, $\tau$, as 
\begin{equation}
 \tau \sim 5 \times 10^{6} {\rm yr}.
\end{equation}
The star formation timescale $t_{\rm SF}$, evaluated at density 
$n_{\rm cl}$, is
\begin{equation}
t_{\rm SF} = A t_{ff} \sim 1.3 \times 10^{7} A \left( \frac{n_{\rm cl}}
{10^{3} {\rm cm}^{-3}} \right)^{-1/2} {\rm yr}.
\end{equation}
Thus, the condition for high SFE 
($\tau > t_{\rm SF}$) in term of $n_{f}$ is 
\begin{equation}
n_{f} > 10 \times \left( \frac{\tau}{5 \times 10^{6} {\rm yr}} \right)^{-2} 
\left( \frac{A}{1}\right)^{2} \left( \frac{C}{800} \right)^{-1}  {\rm cm}^{-3}.
\end{equation}
This criterion is qualitative, since it depends some uncertain factors, 
the same as equation (11). 

For both case, $n_{f} > 10 ~{\rm cm}^{-3}$ is required 
for high SFE and bound cluster formation. 
From Figure 4a we can see that the density $n_{f}$ increases 
as metallicity decreases and also as radiation increases.
The condition $n_{f} > 10 ~{\rm cm}^{-3}$ corresponds to the parameter range 
$Z \la 0.05$ for $X_{0} = 1$ and $X_{0} \ga 8$ for 
$Z=1$.
This means high SFE and efficient bound cluster formation are realized  
in a low-metallicity and/or strong-radiation environment in our model. 
From Figure 4a, we can also see that when $Z \sim 1$, $n_{f}$ is more 
sensitive to $X_{0}$ than $Z$ and when $Z \la 0.1$, $n_{f}$ is more sensitive 
to $Z$ than $X_{0}$.  Thus SFE mainly depends on $X_{0}$ around $Z \sim 1$ and 
on $Z$ below $Z \sim 0.1$.

The qualitative behavior of the above result will be independent of 
the details about evolution of condensations and feedback from star 
formation, since the essential ingredients are only two points, (1) 
low-metallicity and/or strong-radiation fields lead to high breakup density 
of the parent cloud, and (2) high-density environments result in high SFE. 
However, our conclusion may be changed if we consider nonspherical 
evolution of the cloud. 
The dynamical timescale of an initially nonspherical cloud or a cloud with 
rotation is similar to that of a spherical cloud at first, e.g., 
$t_{\rm dyn} \propto n^{-1/2}$, where $n$ is the density of the cloud.
But later disklike collapse will follow 
and the dynamical timescale becomes $t_{\rm dyn} \propto n^{-1}$ 
(Susa, Uehara, $\&$ Nishi 1996). 
Then the ratio of the cooling time $t_{\rm cool} \propto n^{-1} Z^{-1}$ 
to the dynamical time is   
\begin{equation}
\frac{t_{\rm cool}}{t_{\rm dyn}} \propto n^{-1/2} ~Z^{-1},
\end{equation}
for initial spherical-like phase and
\begin{equation}
\frac{t_{\rm cool} }{t_{\rm dyn}} \propto Z^{-1},
\end{equation}
for later disklike phase.
When the cloud becomes thermally unstable, we can estimate whether or not 
condensations form, using the value of $t_{\rm cool}/t_{\rm dyn}$. 
If this ratio is less than unity, condensation will form, and vice versa.  
In case the cloud becomes unstable when it is still in a spherical-like 
phase, condensations will form regardless of the metallicity $Z$,  
as we calculated in this work. This can be explained since 
the density $n$, when the cloud becomes unstable, increases as $Z$ decreases 
(see Figs. 2 and 3), so that $t_{\rm cool}/t_{\rm dyn}$ does not change so 
much. But, in case the cloud becomes unstable when it is in a disklike 
phase, condensations may not form for the low-metallicity case since 
$t_{\rm cool}/t_{\rm dyn}$ increases as $Z$ decreases. In this case 
our discussion in this section is not applicable.
To summarize, in the presence of rotation or deviation from spherical 
symmetry, if the cloud becomes thermally unstable in the early phase of 
collapse, condensations will form and our conclusion in this section is 
adequate. In contrast, if the cloud becomes thermally unstable in the late 
phase of collapse, condensations may not form throughout the collapse and 
another discussion may be required. 

Finally, we present some applications. 
A low-metallicity environment is realized in dwarf galaxies 
and in the early stage of our Galaxy, for example.
And strong-radiation fields are realized in starburst galaxies 
and probably in the early stage of our Galaxy, where young bulge stars 
are expected to provide strong radiation. 
Thus, according to the result, young globular clusters in dwarf galaxies and 
starburst galaxies may be produced through the mechanism we show in this work.
Also, the disk population of globular clusters in our Galaxy (Zinn 1985) may  
be produced through the same mechanism. 

\section{Summary}

In this paper we perform linear analysis of thermal instability in a 
collapsing warm gas cloud and study the effect of metallicity and 
radiation flux. 
For the calculation, we adopt a one-zone approximation and the cloud 
contraction is simplified as free-fall collapse. 
When the cloud density reaches critical value $n_{f}$, the cloud fragments 
into cool dense condensations via thermal instability. 
From our calculation, the critical density $n_{f}$ increases as metallicity 
decreases, and also as radiation increases. 
Condensations collide with each other and self-gravitating clumps  
will be produced when the mean cloud density becomes 
sufficiently high; then stars will form.
Expansion of the \ion{H}{2} region around the massive star and 
supernova explosions will blow off surrounding gas and stop star formation 
process. When the mean density at the time of star formation is high, 
high virial velocity prevents expansion of the \ion{H}{2} region. 
Also, in such high-density environments, the star formation timescale is 
shorter than the lifetime of a massive star. Then the gas in 
cluster-forming region will be converted into stars efficiently, 
before the gas is dispersed by expanding \ion{H}{2} region or supernova 
explosions. 
In our calculation, such high density is realized in the contracting 
low-metallicity gas. And if the formation of a contracting gas cloud is 
possible, a strong-radiation environment is another candidate. 
Thus, it is suggested that high star formation efficiency and bound cluster 
formation are likely achieved in low-metallicity and/or strong-radiation 
environments.
Such environments exist in dwarf galaxies, the early stage of our Galaxy and 
starburst galaxies. 
According to the result, globular clusters in these galaxies 
may be produced through the mechanism we show in this paper.

\acknowledgments

We woud like to thank H. Sato for valuable comments and H. Kamaya for useful 
discssion. 
This work is supported in part by the Japanese Grant-in-Aid for Scientific 
Research on Priority Areas (No. 10147105) of the Ministry of Education, 
Science, Sports and Culture of Japan.

\end{document}